\documentclass{iaus}

\usepackage{graphicx}
\usepackage{epstopdf}

\def\lesssim{\mathrel{\rlap{\lower4pt\hbox{\hskip1pt$\sim$}}}<}
\def\gtrsim{\mathrel{\rlap{\lower4pt\hbox{\hskip1pt$\sim$}}}>}

\title[Dark Stars] {Dark Stars: Dark matter in 
the first stars leads to a new phase of stellar evolution}

\author[Katherine Freese et al.]
{Katherine Freese$^1$, Douglas Spolyar$^2$, Anthony Aguirre$^2$, 
Peter Bodenheimer$^3$, Paolo Gondolo$^4$, J.A. Sellwood$^5$, and
Naoki Yoshida$^6$}

\affiliation{$^1$Michigan Center for Theoretical Physics, University
of Michigan, Ann Arbor, MI 48109, USA \\ email: {\tt ktfreese@umich.edu}
\\[\affilskip]
$^2$Dept. of Physics, University of California, Santa Cruz, CA 95064, USA 
\\ email: {\tt dspolyar@physics.ucsc.edu, aguirre@scipp.ucsc.edu}
\\[\affilskip]
$^3$Dept. of Astronomy, University of California, Santa Cruz, CA 95064, USA 
\\ email: {\tt peter@ucolick.org}
\\[\affilskip]
$^4$Physics Dept., University of Utah, Salt Lake City, UT 84112, USA \\
email: {\tt paolo@physics.utah.edu}
\\[\affilskip]
$^5$Dept. of Physics and Astronomy, Rutgers Univ., Piscataway, NJ 08854, USA\\
email: {\tt sellwood@physics.rutgers.edu}
\\[\affilskip]
$^6$Inst. for the Physics and Math. of the Universe, 
Univ. of Tokyo, Kashiwa, Chiba, 
Japan \\
email: {\tt nyoshida@a.phys.nagoya-u-ac.jp}}

\pubyear{2008}
\volume{255}
\pagerange{}

\jname{Low-Metallicity Star Formation: From the First Stars to Dwarf Galaxies}
\editors{L.K. Hunt, S. Madden \& R. Schneider, eds.}
\begin{document}

\maketitle
\begin{abstract} 
The first phase of stellar evolution in the history of the universe
may be Dark Stars, powered by dark matter heating rather than by fusion.
Weakly interacting massive particles, which are their own
antiparticles, can annihilate and provide an important heat source for
the first stars in the the universe.  This talk presents the story of 
these Dark Stars. We make predictions that the first stars are
very massive ($\sim 800 M_\odot$), cool (6000 K), bright ($\sim 10^6 L_\odot$), 
long-lived ($\sim 10^6$ years), and probable precursors to (otherwise
unexplained) supermassive black holes. 
Later, once the initial DM fuel runs out
and fusion sets in, DM annihilation can predominate again if the scattering
cross section is strong enough, so that a Dark Star is born again.
\end{abstract}

\firstsection

\section{Introduction}

In October 2006, as guests of the Galileo Galilei Institute in Florence,
two of us began a new line of research: the effect
of Dark Matter particles on the very first stars to form in the universe.
We found a new phase of stellar evolution: the first stars to form
in the universe may be ``Dark Stars:''
dark matter powered rather than fusion powered.
We first reported on this work in a paper 
(\cite[Spolyar, Freese, \& Gondolo 2008]{SpolyarFreeseGondolo08})
submitted to the arxiv in April
2007 (hereafter Paper I). When we presented this work at 
The First Stars conference in Santa Fe soon after 
(\cite[Freese,Gondolo \& Spolyar 2008]{FreeseGondoloSpolyar08}), many
questions were raised, which we have addressed in the subsequent year.  
In this talk I review the basic ideas
as well as report on the followup work we performed over the 
past year.

The Dark Matter particles considered here are Weakly Interacting
Massive Particles (WIMPs) (such as the Lightest Supersymmetric Particle),
which are one of the major motivations for building the Large
Hadron Collider at CERN that will begin taking data very soon.
These particles are their own antiparticles; they annihilate among
themselves in the early universe, leaving the correct relic 
density today to explain the dark matter in the universe.  These particles will 
similarly annihilate wherever the DM density is high.
The first stars are particularly good sites for annihilation
because they form at high redshifts (density scales as $(1+z)^3$) and
in the high density centers of DM haloes.  
The first stars form at redshifts $z \sim 10-50$ 
in dark matter (DM) haloes
of $10^6 M_\odot$ (for reviews see e.g. 
\cite[Ripamonti \& Abel 2005]{RipamontiAbel05},
\cite[Barkana \& Loeb 2001]{BarkanaLoeb01},
\cite[Bromm \& Larson 2003]{BrommLarson03}; see also
\cite[Yoshida et al. 2006]{Yoshida_etal06}.)
One star is thought to form inside one such
DM halo. 

As our canonical values, we will use the standard annihilation 
cross section,
$\langle \sigma v \rangle = 3 \times 10^{-26} {\rm cm^3/sec}$,
and $m_\chi = 100$ GeV for the particle mass;
but we will also consider a broader range of 
masses and cross-sections.
Paper I  found that
DM annihilation provides a powerful heat source in the first stars, a
source so intense that its heating overwhelms all cooling mechanisms;
subsequent work has found that the heating dominates over fusion as well
once it becomes important at later stages (see below). 
Paper I (\cite[Spolyar, Freese, \& Gondolo 2008]{SpolyarFreeseGondolo08})
suggested that the
very first stellar objects might be ``Dark Stars,'' a new phase of
stellar evolution in which the DM -- while only a negligible fraction
of the star's mass -- provides the power source for the star through
DM annihilation.

\begin{figure}[t]
\centerline{\includegraphics[width=0.5\textwidth]{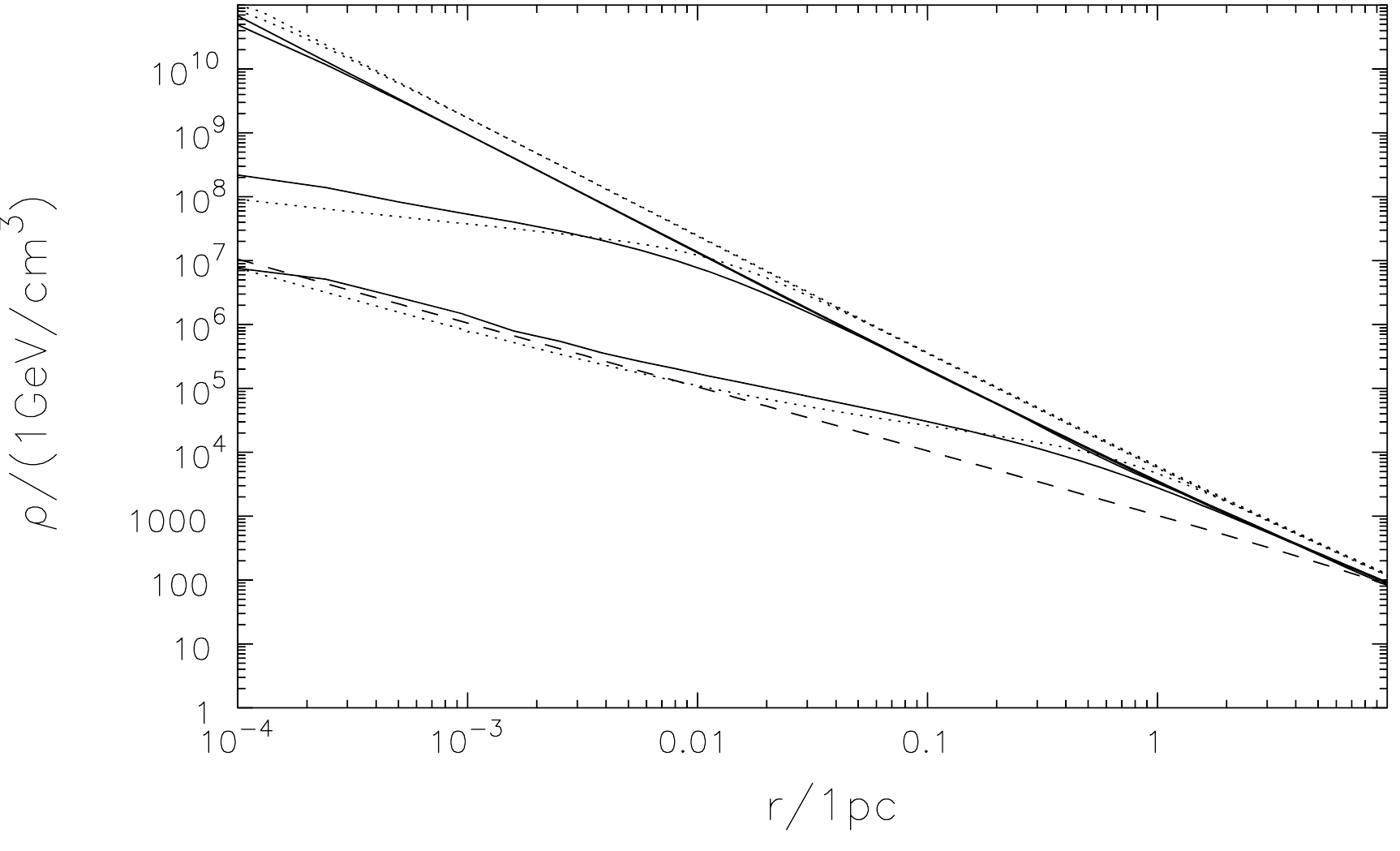}}
\caption{Adiabatically contracted DM profiles in the first
protostars for an initial NFW
profile (dashed line) using (a) the Blumenthal method (dotted lines)
and (b) an exact calculation using
Young's method (solid lines), for $M_{\rm vir}=5 \times 10^7
M_\odot$, $c=2$, and $z=19$.  The four sets of curves 
correspond to a baryonic core density of $10^4, 10^8, 10^{13},$ and
$10^{16}{\rm cm}^{-3}$.  The two different
approaches to obtaining the DM densities find values that differ by
less than a factor of two.
}
\end{figure}

\section{Three Criteria}

Paper I (\cite[Spolyar, Freese, \& Gondolo 2008]{SpolyarFreeseGondolo08})
outlined the three key ingredients for Dark Stars:
1) high dark matter densities, 2) the annihilation products get
stuck inside the star, and 3) DM heating wins over other cooling
or heating mechanisms.  These same ingredients are required throughout
the evolution of the dark stars, whether during the protostellar
phase or during the main sequence phase.

{\bf First criterion: High Dark Matter density inside the star.}
To find the DM density profile, we start with an 
overdense region of $\sim 10^6 M_\odot$
with an NFW (\cite[Navarro, Frenk \& White 1996]{NavarroFrenkWhite96}) 
profile for both DM and gas, where the gas contribution is
15\% of that of the DM. Originally we used adiabatic contraction ($M(r)r$ =
constant) (\cite[Blumenthal et al. 1985]{Blumenthal_etal85}) and matched
onto the baryon density profiles given by 
\cite[Abel, Bryan \& Norman 2002]{AbelBryanNorman02}
and \cite[Gao et al. 2007]{Gao_etal07} to
obtain DM profiles.  This method is overly simplified:
it considers only circular orbits of the DM particles. 
Our original DM profile matched
that obtained numerically in
\cite[Abel, Bryan, \& Norman 2002]{AbelBryanNorman02}
with $\rho_\chi \propto r^{-1.9}$, for both their 
earliest and latest profiles; see also 
\cite[Natarajan, Tan, \& O'Shea 2008]{NatarajanTanO'Shea08} for a recent
discussion.  Subsequent to our original work, 
we have done an exact calculation (which includes
radial orbits) 
(\cite[Freese, Gondolo, Sellwood \& 
Spolyar 2008]{FreeseGondoloSellwoodSpolyar08})
and found that our original results were remarkably 
accurate, to within a factor of two.  Our resultant 
DM profiles are shown in Fig.~1.
At later stages, we also consider possible further enhancements
due to capture of DM into the star (discussed below).

{\bf Second Criterion: Dark Matter Annihilation Products get stuck inside
the star}.
WIMP annihilation produces energy at a rate per unit volume $Q_{\rm
ann} = \langle \sigma v \rangle \rho_\chi^2/m_\chi \linebreak \simeq  1.2
\times  10^{-29} {\rm erg/cm^3/s} \,\,\, (\langle \sigma v \rangle / (3
\times 10^{-26} {\rm cm^3/s}))  (n/{\rm cm^{-3}})^{1.6} (m_\chi/(100
{\rm GeV}))^{-1}$. In the early stages of Pop III star formation, when
the gas density is low, most of this energy is radiated away 
(\cite[Ripamonti Mapelli \& Ferrara 2006]{RipamontiMapelliFerrara06}). However,
as the gas collapses and its density increases, a substantial fraction
$f_Q$ of the annihilation energy is deposited into the gas, heating it
up at a rate $f_Q Q_{\rm ann}$ per unit volume.   While neutrinos
escape from the cloud without depositing an appreciable amount of energy,
electrons and photons  can transmit energy to the core.  
We have computed estimates of this fraction $f_Q$ as the core becomes more
dense. Once $n\sim 10^{11} {\rm cm}^{-3}$ (for 100 GeV WIMPs),  e$^-$ and 
photons are trapped and we can take $f_Q \sim 2/3$.

{\bf Third Criterion: DM Heating is the dominant heating/cooling mechanism
in the star}.
We find that, for WIMP mass
$m_\chi = 100$GeV (1 GeV), a crucial transition takes place when
the gas density reaches $n> 10^{13} {\rm cm}^{-3}$ ($n>10^9
{\rm cm}^{-3}$).  Above this density, DM heating dominates over
all relevant cooling mechanisms, the most important being
H$_2$ cooling
(\cite[Hollenbach \& McKee 1979]{HollenbachMcKee79}).

Figure 2 shows evolutionary tracks of the protostar
in the temperature-density phase plane with DM heating included
(\cite[Yoshida et al. 2008]{Yoshida_etal08}),
for two DM particle masses (10 GeV and 100 GeV).  
Moving to the right on this
plot is equivalent to moving forward in time.  Once the
black dots are reached, DM heating dominates over cooling inside the star,
and the Dark Star phase begins.
The protostellar core is 
prevented from cooling and collapsing further.  
The size of the core at this point 
is $\sim 17$ A.U. and its mass is $\sim 0.6 M_\odot$ 
for 100 GeV mass WIMPs.
A new type of object is created, a Dark Star
supported by DM annihilation rather than fusion.  

\begin{figure}[t]
\centerline{\includegraphics[width=0.5\textwidth]{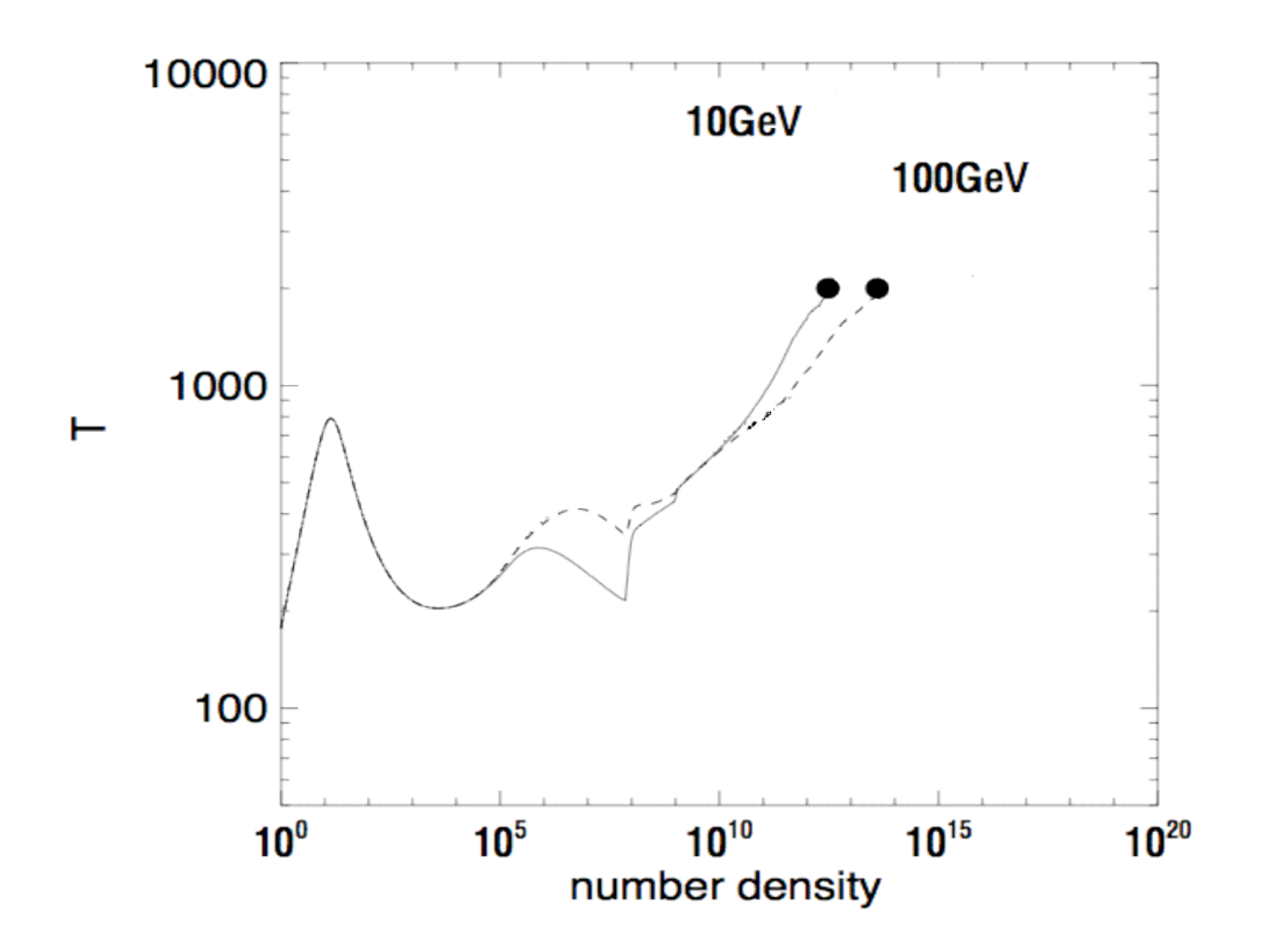}}
\caption{ Temperature (in degrees K) as a function of hydrogen density 
(in cm$^{-3}$) for the first protostars, with DM annihilation included,
for two different DM particle masses (10 GeV and 100 GeV).
Moving to the right in the figure corresponds to moving forward in time.
Once the ``dots'' are reached, DM annihilation wins over H2 cooling,
and a Dark Star is created.
}
\end{figure}

\section{Building up the Mass}

Recently, we have found the stellar structure
of the dark stars (hereafter DS) 
(\cite[Freese, Bodenheimer, Spolyar, \& Gondolo 2008]{FreeseBodenheimerSpolyarGondolo08}).  
Though they form with the properties just mentioned,
they continue to accrete mass from the surrounding medium. 
In our paper we build up the DS
mass as it grows from $\sim 1 M_\odot$
to $\sim 1000 M_\odot$.  
As the mass increases, the DS contracts and the DM
density increases until the DM heating matches its radiated luminosity.  We find
polytropic solutions for dark stars in hydrostatic and thermal
equilibrium.  We start
with a few $M_\odot$ DS and find an equilibrium solution.  Then we build up
the DS by accreting $1 M_\odot$ at a time with an accretion rate
of $2 \times 10^{-3} M_\odot$/yr, always finding equilibrium
solutions.  We find that initially
the DS are in convective equilibrium; from $(100-400)
M_\odot$ there is a transition to radiative; and heavier
DS are radiative.  As the DS grows, it pulls in more
DM, which then annihilates.  We continue this process until
the DM fuel runs out at $M_{DS} \sim 800 M_\odot$ (for 100 GeV WIMPs). 
Figure 3 shows the stellar structure. One can see
``the power of darkness:'' although the DM constitutes a tiny fraction
($<10^{-3}$) of the mass of the DS, it can power the star. The reason
is that WIMP annihilation is a very efficient power source:
2/3 of the initial energy of the WIMPs is converted into useful
energy for the star, whereas only 1\% of baryonic rest mass energy
is useful to a star via fusion.

\begin{figure}[t]
\centerline{\includegraphics[width=0.5\textwidth]{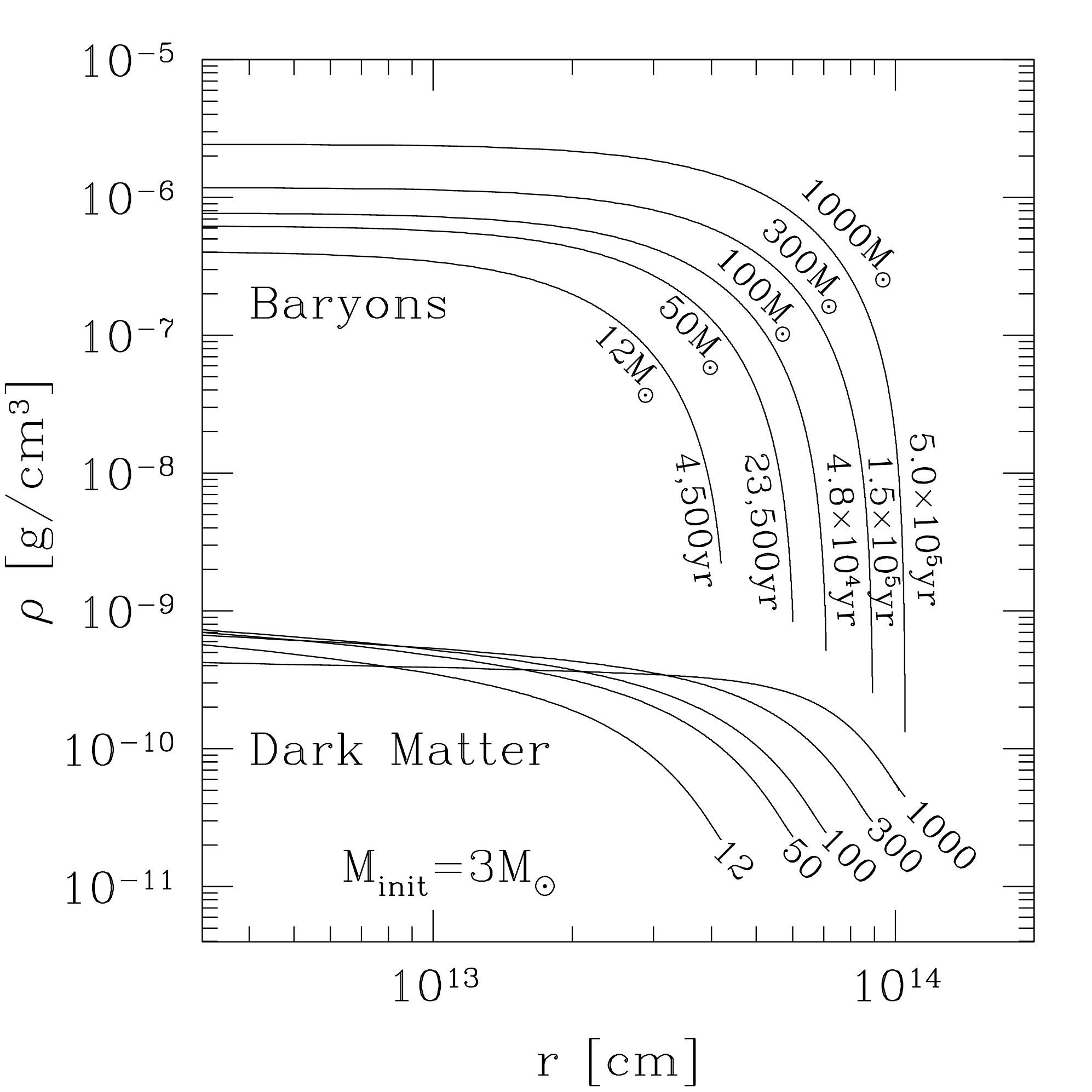}}
\caption{ Evolution of a dark star (n=1.5) as mass is accreted onto the initial
protostellar core of 3 M$_\odot$.  The set of upper
(lower) curves correspond to the baryonic (DM) density profile at different
masses and times.  Note that DM constitutes $<10^{-3}$ of the mass of the DS.
\vspace{-0.5\baselineskip}
}
\end{figure}

\section{Results and Predictions}

 Our final result 
(\cite[Freese, Bodenheimer, Spolyar, \& Gondolo 2008]{FreeseBodenheimerSpolyarGondolo08}), 
is {\bf very large first stars!}
E.g., for 100 GeV WIMPs, the first stars have $M_{DS} = 800 M_\odot$.
Once the DM fuel runs out inside the DS, the star contracts until
it reaches $10^8$K and fusion sets in.
A possible end result of stellar evolution will be {\bf large black holes}. 
The Pair Instability
SN (\cite[Heger \& Woosley 2002]{HegerWoosley02})
that would be produced from 140-260 $M_\odot$ stars (and whose
chemical imprint is not seen) would not be as abundant.  Indeed 
this process may help to explain the supermassive black
holes that have been found at high redshift ($10^9 M_\odot$ BH at
z=6) and are, as yet, unexplained 
(\cite[Li et al. 2006]{Li_etal06};
\cite[Pelupessy et al. 2007]{Pelupessy_etal07}). 

The lifetime of this new DM powered phase of stellar evolution, 
prior to the onset of fusion,
is $\sim 10^6$ years. The stars are very bright,
$\sim 10^6 L_\odot$, and relatively cool, (6000-10,000)K (as opposed
to standard Pop III stars whose surface temperatures exceed $30,000K$).  
Reionization
during this period is likely to be slowed down, as these stars can
heat the surroundings but not ionize them.  
One can thus
hope to find DS and differentiate them from standard Pop III stars;
perhaps some even still exist to low redshifts.

\section{Later stages: Capture}

There is another possible source of DM in the first stars: capture
of DM particles from the ambient medium. Whereas capture is negligible
during the pre-mainsequence phase, once fusion sets in it can be important,
depending on the value of the scattering cross section of DM with the gas.
Two simultaneous papers
(\cite[Freese, Spolyar, \& Aguirre 2008]{FreeseSpolyarAguirre08},
\cite[Iocco 2008]{Iocco08})
found the same basic idea: the DM luminosity from captured WIMPs
can be larger than fusion for the DS. Two uncertainties exist here:
the scattering cross section, and the amount of DM in the ambient medium
to capture from\footnote{Unlike the annihilation cross section, 
which is set by the relic
density, scattering is to some extent a free parameter set only
by bounds from direct detection experiments.}.
DS studies including capture
have assumed the maximal scattering cross sections allowed by
experimental bounds and ambient DM densities that are never depleted.
With these assumptions, 
DS evolution models with DM heating after the onset of fusion
have now been studied in several papers
(\cite[Iocco et al. 2008]{Iocco_etal08},
\cite[Taoso et al. 2008]{Taoso_etal08},
\cite[Yoon et al 2008]{Yoon_etal08})
which were posted during
the conference and will be discussed
in the talk by Fabio Iocco.  We have been pursuing similar research
with Alex Heger on DS  
evolution after the onset of fusion.  

\section{Conclusion}

The line of research we began in Florence almost two
years ago is reaching a very fruitful stage of development.
Dark matter can play a crucial role in the first stars.  The first
stars to form in the universe may be Dark Stars: powered by DM heating
rather than by fusion.  Our work indicates that they may 
be very large ($850 M_\odot$ for 100 GeV mass WIMPs).
The connections between particle physics
and astrophysics are ever growing!

\end{document}